\documentclass{article}

\usepackage{arxiv}

\usepackage[utf8]{inputenc} 
\usepackage[T1]{fontenc}    
\usepackage{hyperref}       
\usepackage{url}            
\usepackage{booktabs}       
\usepackage{amsfonts}       
\usepackage{nicefrac}       
\usepackage{microtype}      
\usepackage{lipsum}		
\usepackage{graphicx}
\usepackage{natbib}
\usepackage{doi}
\usepackage{siunitx}
\usepackage{amsmath,amssymb,amsfonts}
\usepackage{mathtools} 
\newcommand{\centered}[1]{\begin{tabular}{l} #1 \end{tabular}}
\usepackage{amssymb}
\usepackage{pifont}
\newcommand{\cmark}{$\checkmark$}
\newcommand{\xmark}{$\times$}

\title{Neuromorphic-P\si{^2}M: \underline{P}rocessing-in-\underline{P}ixel-in-\underline{M}emory Paradigm for Neuromorphic Image Sensors}

\author{Md Abdullah-Al Kaiser\thanks{These authors contributed equally to this work.} \\
	Department of Electrical and Computer Engineering\\
	University of Southern California\\
	Los Angeles, CA 90089 \\
	\texttt{mdabdull@usc.edu} \\
	\And
	Gourav Datta$^*$ \\
	Department of Electrical and Computer Engineering\\
	University of Southern California\\
	Los Angeles, CA 90089 \\
	\texttt{gdatta@usc.edu} \\
	\And
	Zixu Wang \\
	Department of Electrical and Computer Engineering\\
	University of Southern California\\
	Los Angeles, CA 90089 \\
	\texttt{zixuwang@usc.edu} \\
 	\And
	Ajey P. Jacob \\
	Information Sciences Institute\\
	University of Southern California\\
	Los Angeles, CA 90292 \\
	\texttt{ajey@isi.edu} \\
 	\And
	Peter A. Beerel \\
	Department of Electrical and Computer Engineering\\
	University of Southern California\\
	Los Angeles, CA 90089 \\
	\texttt{pabeerel@usc.edu} \\
  	\And
	Akhilesh R. Jaiswal \\
	Department of Electrical and Computer Engineering\\
	University of Southern California\\
	Los Angeles, CA 90089 \\
	\texttt{akhilesh@usc.edu} \\
}

\renewcommand{\shorttitle}{Neuromorphic-P\si{^2}M}

\hypersetup{
pdftitle={A template for the arxiv style},
pdfsubject={q-bio.NC, q-bio.QM},
pdfauthor={David S.~Hippocampus, Elias D.~Striatum},
pdfkeywords={First keyword, Second keyword, More},
}

\begin{document}
\maketitle

\begin{abstract}
Edge devices equipped with computer vision must deal with vast amounts of sensory data with limited computing resources. Hence, researchers have been exploring different energy-efficient solutions such as near-sensor processing, in-sensor processing, and in-pixel processing, bringing the computation closer to the sensor. In particular, in-pixel processing embeds the computation capabilities inside the pixel array and achieves high energy efficiency by generating low-level features instead of the raw data stream from CMOS image sensors. Many different in-pixel processing techniques and approaches have been demonstrated on conventional frame-based CMOS imagers, however, the processing-in-pixel approach for neuromorphic vision sensors has not been explored so far. 
In this work, we for the first time, propose an asynchronous non-von-Neumann analog processing-in-pixel paradigm to perform convolution operations by integrating in-situ multi-bit multi-channel convolution inside the pixel array performing analog multiply and accumulate (MAC) operations that consume significantly less energy than their digital MAC alternative. To make this approach viable, we  incorporate the circuit's non-ideality, leakage, and process variations into a novel hardware-algorithm co-design framework that leverages extensive HSpice simulations of our proposed circuit using the GF22nm FD-SOI technology node. 
We verified our framework on state-of-the-art neuromorphic vision sensor datasets and show that our solution consumes $\sim2{\times}$ lower backend-processor energy while maintaining almost similar front-end (sensor) energy on the IBM DVS128-Gesture dataset than the state-of-the-art while maintaining a high test accuracy of $88.36\%$. 
\end{abstract}


\section{Introduction}
Today's widespread video acquisition and interpretation applications (e.g., autonomous driving \cite{auto_driving}, surveillance \cite{surveillance}, object detection \cite{vid_obg_detect}, object tracking \cite{obj_track}, and anomaly detection \cite{anomaly_detect}) are fueled by CMOS image sensors (CIS) and deep learning algorithms. However, these computer vision systems suffer from energy efficiency and throughput bottlenecks \cite{system_bottleneck} that stem from the transmission of a high volume of data between the sensors at the edge and processors in the cloud. For example, smart glasses (e.g., Meta AR/VR glasses, google classes, etc.) drain the battery within 2-3 hours when used for intensive computer vision tasks \cite{google_glass}. Although there are significant technological and system-level advancements in both CMOS imagers \cite{LPCIS_survey} and deep neural networks \cite{DNN_survey}, the underlying energy inefficiency arises due to the physical separation of sensory and processing hardware. Hence, developing novel energy-efficient hardware for resource-constrained computer vision applications has attracted significant attention in the research community.

Many researchers implement the first few computation tasks of machine vision applications close to the sensor to reduce the energy consumption of massive data transfer \cite{near_in_sensor_survey}. These approaches can be categorized into three types (1) near-sensor processing, (2) in-sensor processing, and (3) in-pixel processing. In the near-sensor processing approach, a digital signal processor or machine learning accelerator is placed close to the sensor chip. In \cite{near_sensor_ARVR}, a dedicated near-sensor processor led to a 64.6\% drop in inference energy for MobileNetV3.  In \cite{near_sensor_3D_sony}, a 3D stacked system consisting of a CNN inference processor and a back-side illuminated CMOS image sensor demonstrated an energy efficiency of 4.97 TOPS/W. Near-sensor computing can improve energy efficiency by reducing the data transfer cost between the sensor chip and the cloud/edge processor, however, the data traffic between the sensor and near-sensor processor still consumes significant amounts of energy.

In contrast, the in-sensor approach utilizes an analog or digital signal processor at the periphery of the sensor chip. For instance, RedEye \cite{redeye} uses analog convolution processing before the sensor's analog-to-digital conversion (ADC) blocks to obtain a 5.5\si{\times} reduction in sensor energy. Moreover, a mixed-mode in-sensor tiny convolution neural network (CNN) \cite{mixed_mode_ivs} yielded a significant reduction in bandwidth and in particular reduced the power consumption associated with the ADC. To fully remove the ADC energy overhead, \cite{analog_bnn_swcap} processed the raw analog data from the CMOS image sensor using an on-chip completely analog binary neural network (BNN) that leverages switched capacitors. Using energy-efficient analog computing was also explored in \cite{analog_bnn} which proposes a novel current-mode analog low-precision BNN.  Furthermore, SleepSpotter \cite{sleepspotter} implemented energy-efficient current-domain on-chip MAC operations.  Nevertheless, this solution still requires the potentially-compressed raw analog data to be streamed through column-parallel bitlines from the sensor nodes to the peripheral processing networks. In general, these in-sensor approaches significantly reduce the energy overhead of the analog-to-digital converters, however, they still suffer from the data transfer bottleneck between the sensor and peripheral logic.

On the other hand, the in-pixel processing approach integrates computation capabilities inside the pixel array to enable early processing and minimize the subsequent data transmission. For instance, a low-voltage in-pixel convolution operation has been proposed in \cite{pwm_idac_weight} that utilizes a current-based digital-to-analog converter (DAC) to implement weights and pulse-width-modulated (PWM) pixels. Moreover, a single instruction multiple data (SIMD) pixel processor array (PPA) \cite{simd_ppa} can perform parallel convolution operations within the pixel array by storing the weights of the convolution filters in registers within the in-pixel processing elements. In addition, the direct utilization of the photodetector current to compute the binary convolution can yield 11.49 TOPS/W energy efficiency \cite{Iph_sw}. Furthermore, \cite{senputing} performs classification tasks on the MNIST dataset by generating the in-pixel MAC results of the first BNN layer and exhibits 17.3 TOPS/W energy efficiency. In addition, a processing-in-pixel-in-memory paradigm for CIS reported an 11\si{\times} energy-delay product (EDP) improvement on the Visual Wake Words (VWW) dataset \cite{aps_p2m}. Follow-up works by the same authors have demonstrated 5.26\si{\times} and 3.14\si{\times} reduction in energy consumption on hyperspectral image recognition \cite{p2m_hsi} and multi-object tracking in the wild \cite{p2mdetrack}, respectively. In summary, due to the embedded pixel-level processing elements, the in-pixel processing approach can outperform energy and throughput compared to in-sensor and near-sensor processing solutions.

Most of the research works on different energy-efficient CIS approaches (near-sensor, in-sensor, and in-pixel processing) are focused on conventional frame-based imagers. However, many researchers are now exploring the use of event-driven neuromorphic cameras or dynamic vision sensors (DVS) \cite{DVS_ref1}, \cite{DVS_ref2} for different neural network applications, including autonomous driving \cite{DVS_auto_driving}, steering angle prediction \cite{DVS_steering}, optical flow estimation \cite{DVS_opt_flow}, pose re-localization \cite{DVS_pose}, and lane marker extraction \cite{DVS_lane}, due to their energy, latency, and throughput advantages over traditional CMOS imagers. 
The DVS pixel generates event spikes based on the change in light intensity instead of sensing the absolute pixel-level illumination in conventional CMOS imagers. Thus DVS pixels filter out the redundant information from a visual scene and produce sparse asynchronous events. These sparse events are communicated off-chip using the address event link protocol \cite{aer_link}. By avoiding the analog-to-digital conversion of the absolute pixel intensity and frame-based sensing method, DVS exhibits higher energy efficiency, lower latency, and higher throughput than frame-based alternatives. Moreover, the dynamic range of the DVS pixel is higher than the conventional CMOS imagers, hence, the DVS camera can adapt to the illumination level of the scene due to its logarithmic receptor. These advantages motivate a paradigm shift towards neuromorphic vision sensors for vision-based applications.
 
These DVS cameras are often coupled with spiking convolution neural networks (CNN) that natively accept asynchronous input events. Traditionally, time is decomposed into windows and the number of spikes that occur in each time window is accumulated independently for each pixel creating multi-bit inputs to a spiking CNN. The first spiking CNN layer thus consists of digital MAC operations (not accumulations because the input is multi-bit instead of binary) unlike the subsequent spiking CNN layers that consist of more energy-efficient accumulations that operate on spikes \citep{datta_date,datta_frontiers}.  
To improve the energy efficiency of such a DVS system, this paper explores in-pixel processing by performing MAC operations in the analog domain within the pixel array. In particular, we have developed a novel energy-efficient neuromorphic processing-in-pixel-in-memory (P\si{^2}M) computing paradigm in which we implement the first spiking CNN layer using embedded transistors that model the multi-bit multi-channel weights and enable massively parallel in-pixel spatio-temporal MAC operations. Because the DVS event spikes are asynchronous in nature, we perform the multiply operation by accumulating the associated weight each time a pixel event occurs. We threshold the accumulated value at the end of each time window to produce a binary output activation and reset the accumulator in preparation for the next time window. To support the presence of multiple input filters that operate on individual pixels, we parallelize this operation and operate on all channels (and all pixels) simultaneously. This charge-based in-pixel analog MAC operation exhibits higher energy efficiency than its digital off-chip counterpart. Moreover, the sparse binary output activations are communicated utilizing a modified address-event representation (AER) protocol, hence, preserving the energy benefit of the workload sparsity. In addition, we have developed a hardware-algorithm co-design framework incorporating the circuit's non-linearity, process variation, leakage, and area consideration based on the GF22nm FD-SOI technology node. Finally, we have demonstrated the feasibility of our hardware-algorithm framework utilizing state-of-the-art neuromorphic event-driven datasets (e.g., IBM DVS128-Gesture, NMNIST) and evaluated our approach's performance and energy improvement. We incur a ${\sim}$5$\%$ accuracy drop in these datasets because our charge-based P\si{^2}M approach does not capture the conventional notion of membrane potential for the first CNN layer. This lack of membrane potential is due to the limited length of time a passive analog capacitor can effectively store charge without significant leakage. However, this problem can be mitigated using non-volatile memories \cite{NVM_memory} that we plan to explore in our future work.

The key contributions of our work are as follows:
\begin{enumerate}
    \item We propose a novel neuromorphic-processing-in-pixel-in-memory (Neuromorphic-P\si{^2}M) paradigm for neuromorphic image sensors, wherein, multi-bit pixel-embedded weights enable massively parallel spatio-temporal convolution on input events inside the pixel array. 
    \item Moreover, we propose non-von-Neumann charge-based energy-efficient in-pixel asynchronous analog multiplication and accumulation (MAC) units and incorporate the non-idealities and process variations of the analog convolution blocks into our algorithmic framework.
    \item Finally, we develop a hardware-algorithm co-design framework considering hardware constraints (non-linearity, process variations, leakage, area consideration), benchmark the accuracy, and yield a $\sim2\times$ improvement in backend-processor energy consumption on the IBM DVS128-Gesture dataset with a $\sim5\%$ drop in test accuracy.
\end{enumerate}

The remainder of the paper is organized as follows. Section 2 describes the circuit implementation, operation, and manufacturability of our proposed Neuromorphic-P\si{^2}M approach. Section 3 explains our hardware-algorithm co-design approach and hardware constraints on the first layer of the neural network model. Section 4 demonstrates our experimental results on two different event-driven DVS datasets, and evaluates the accuracy and performance metrics. Finally, Section 5 presents the concluding remarks.

\section{P\si{^2}M Circuit Implementation}
\begin{figure}[!t]
\begin{center}
\includegraphics[width=0.9\linewidth]{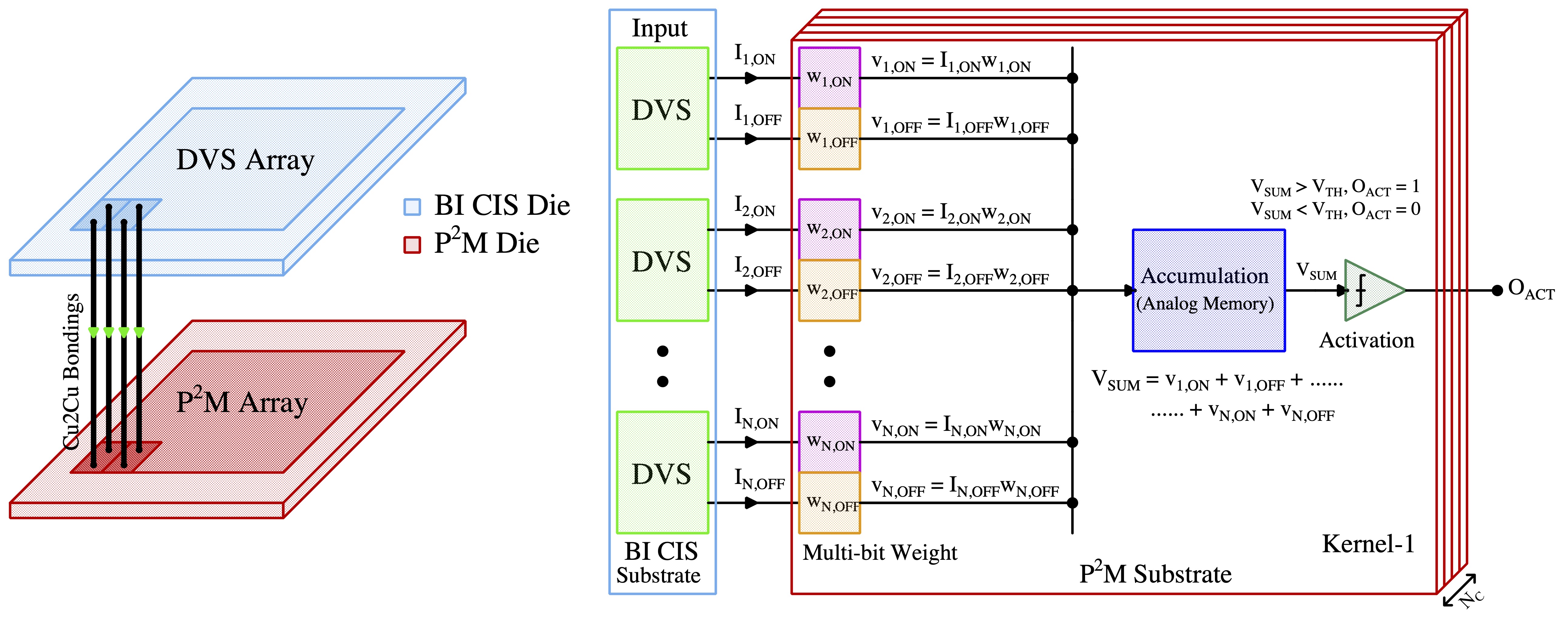}
\end{center}
\caption{The representative 3D chip stack and computing flow diagram of our proposed Neuromorphic P\si{^2}M architecture.}
\label{compute_flow}
\end{figure}

This section presents the critical hardware innovations and implementation of our proposed neuromorphic-P\si{^2}M approach. Figure \ref{compute_flow} illustrates the representative chip stack and computing flow for the first convolution layer utilizing our proposed neuromorphic-P\si{^2}M architecture. The top die consists of DVS pixels and generates ON (OFF) events based on the increase (decrease) in input light contrast level. A DVS pixel consists of a logarithmic receptor, source-follower buffer, capacitive-feedback difference amplifier, and two comparators \cite{DVS_base}, \cite{DVS_ref1}, \cite{DVS_ref2}. The generated events (ON and OFF) per pixel are communicated to the bottom die via pixel-level hybrid Cu-to-Cu interconnects. The bottom die contains the weights and energy-efficient charge-based analog convolution blocks. Each DVS pixel's output channel (ON-channel and OFF-channel) is connected to a transistor in the bottom die that implements a multi-bit weight (e.g., \si{w_{1,ON}}, \si{w_{1,OFF}}, etc.) to perform the multiplication (e.g., \si{I_{1,ON} \times w_{1,ON}}, \si{I_{1,OFF} \times w_{1,OFF}}, etc.) operation. The positive and negative weights are implemented by utilizing the pMOS and nMOS transistors, respectively. Each kernel (corresponding to the filter of the spiking CNN model) accumulates its weighted multiplication of input events on an analog memory (capacitor) asynchronously when an ON or OFF event occurs in the input DVS pixel. As the input spikes are binary, the accumulation voltage either steps up (positive weight) or down (negative weight) by an amount depending on the weight values. The accumulation continues for a fixed time period (simulation time length for each event stream of our neural network model) and after that, the summed voltage is compared with the threshold (using a comparator or skewed inverter) to generate the output activation signal (e.g., \si{O_{ACT}}) of each kernel for the next layer. A similar computing flow is used across the different kernels throughout the sensor array. 

The operations of our proposed neuromorphic-P\si{^2}M can be divided into three phases. These are:-
\begin{enumerate}
    \item Reset Phase: During the reset phase, the accumulation capacitor of each kernel is precharged to 0.5\si{V_{DD}} so that the accumulation voltage can step up or down within the supply rail depending on positive or negative weights, respectively. 
    
    \item Convolution Phase: In the convolution phase, the multi-bit weight-embedded pixels and the accumulation capacitor of each kernel perform multiplication and accumulation (MAC) operations in the analog domain for a fixed period of time. After that, the final accumulated voltage of each kernel is compared with a threshold voltage to (potentially) generate the output activation spike for the next layer.
    
    \item Read Phase: Finally, during the read phase, the output activations of different kernels are sequentially read utilizing the asynchronous Address-Event Representation (AER) read scheme. 
\end{enumerate}

More details on each step including their hardware implementations will be explained below.

\subsection{Multi-bit Weight Embedded Pixels}

\begin{figure}[!t]
\begin{center}
\includegraphics[width=0.8\linewidth]{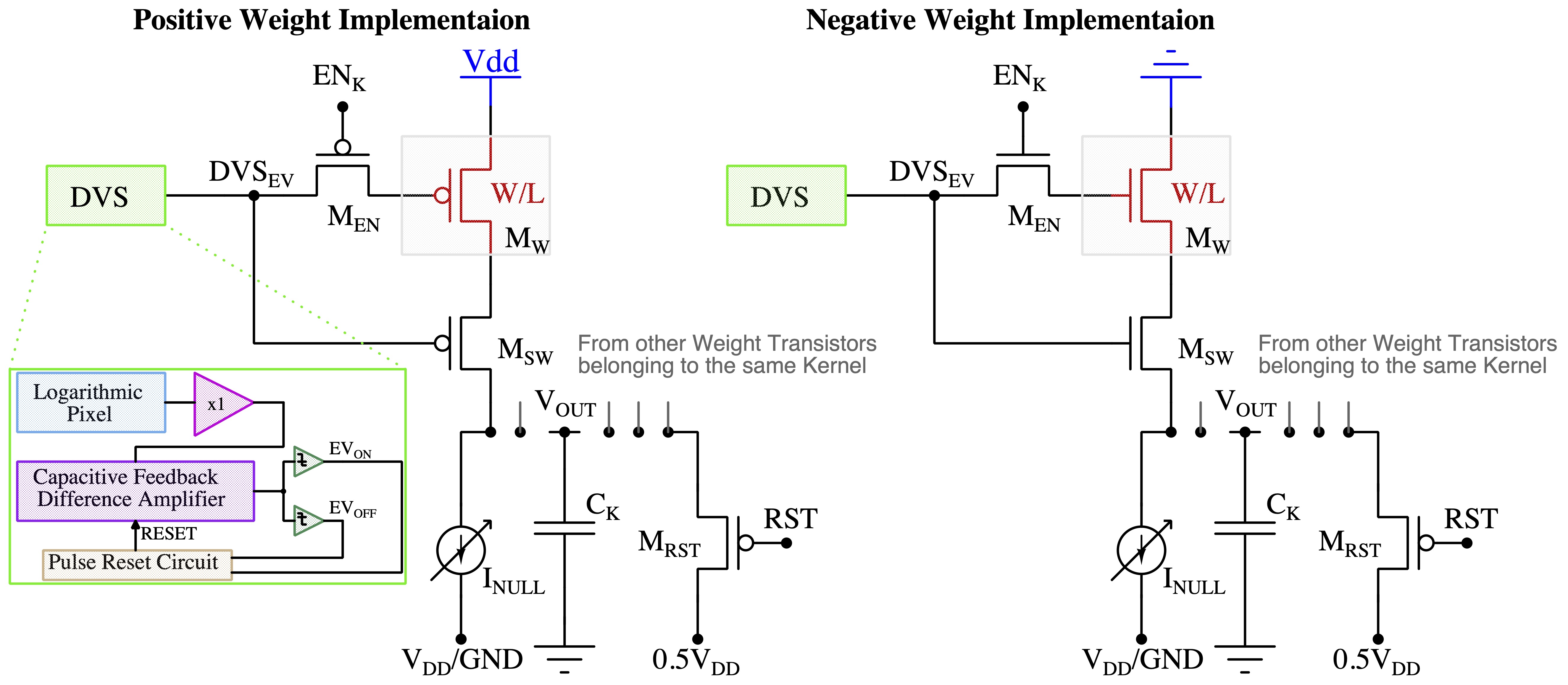}
\end{center}
\caption{Embedded multi-bit positive and negative weight implementation.}
\label{embedded_weight}
\end{figure}

As illustrated in Figure \ref{embedded_weight}, positive and negative weights of the first spiking CNN layer have been implemented by utilizing pMOS and nMOS transistors connected with supply voltages \si{V_{DD}} and ground, respectively.  For a positive (negative) weight, the voltage across the kernel's capacitor (\si{C_K}) charges (discharges) from 0.5\si{V_{DD}} to \si{V_{DD}} (ground) as a function of weight values and the number of input DVS events. The weight values can be tuned by varying the driving strength (\si{\frac{W}{L}} ratio) of the weight transistors (\si{M_W}). A high-\si{V_T} pMOS in positive weight implementation (nMOS in negative weight implementation) (\si{M_{EN}}) is activated during the convolution phase to enable the multiplication and accumulation operations on the kernel's capacitor (\si{C_K}) and remains off during the reset phase. The weight transistor (\si{M_W}) is chosen to have a high-\si{V_T} to limit the charging (for positive weight) or discharging (for negative weight) current to avoid capacitor saturation. Moreover, each DVS pixel includes a delayed self-reset circuit (consisting of a current-starved inverter chain and AND gate) to prevent voltage saturation on the capacitor (\si{C_K}) by limiting the event pulse duration. A switch transistor (\si{M_{SW}}) controlled by the DVS event spike is used to isolate the kernel's capacitor (\si{C_K}) from the weight transistor (\si{M_W}) to reduce the leakage. The switching transistor (\si{M_{SW}}) will be activated only when there are input DVS spikes, hence, ensuring the asynchronous MAC operation on the kernel's capacitor (\si{C_K}). 
Furthermore, to reduce the leakage a kernel-dependent (as leakage is a function of transistor's geometry, hence, leakage amount is dependent on the kernel's weights) current source (\si{I_{NULL}}) is connected with the accumulation capacitor (\si{C_K}) that flows in the opposite direction of the leakage current to nullify the leaky behavior of the capacitor. 
The number of weight transistors associated with a kernel depends on the size of the kernel (e.g., for a kernel size of 3\si{\times}3, there will be a total of 18 weight transistors considering the ON and OFF-channel). For each kernel, the weight transistors are connected to one accumulation capacitor (\si{C_K}). 
Note that the weights cannot be re-programmed after the manufacturing process. However, it is common to use pre-trained weights for the first few layers as low-level feature extractors in modern neural network models \cite{fixed_weight}. Hence, the fixed weights of our proposed architecture do not limit its application for a wide range of machine-vision tasks.

\begin{figure}[!b]
\begin{center}
\includegraphics[width=0.8\linewidth]{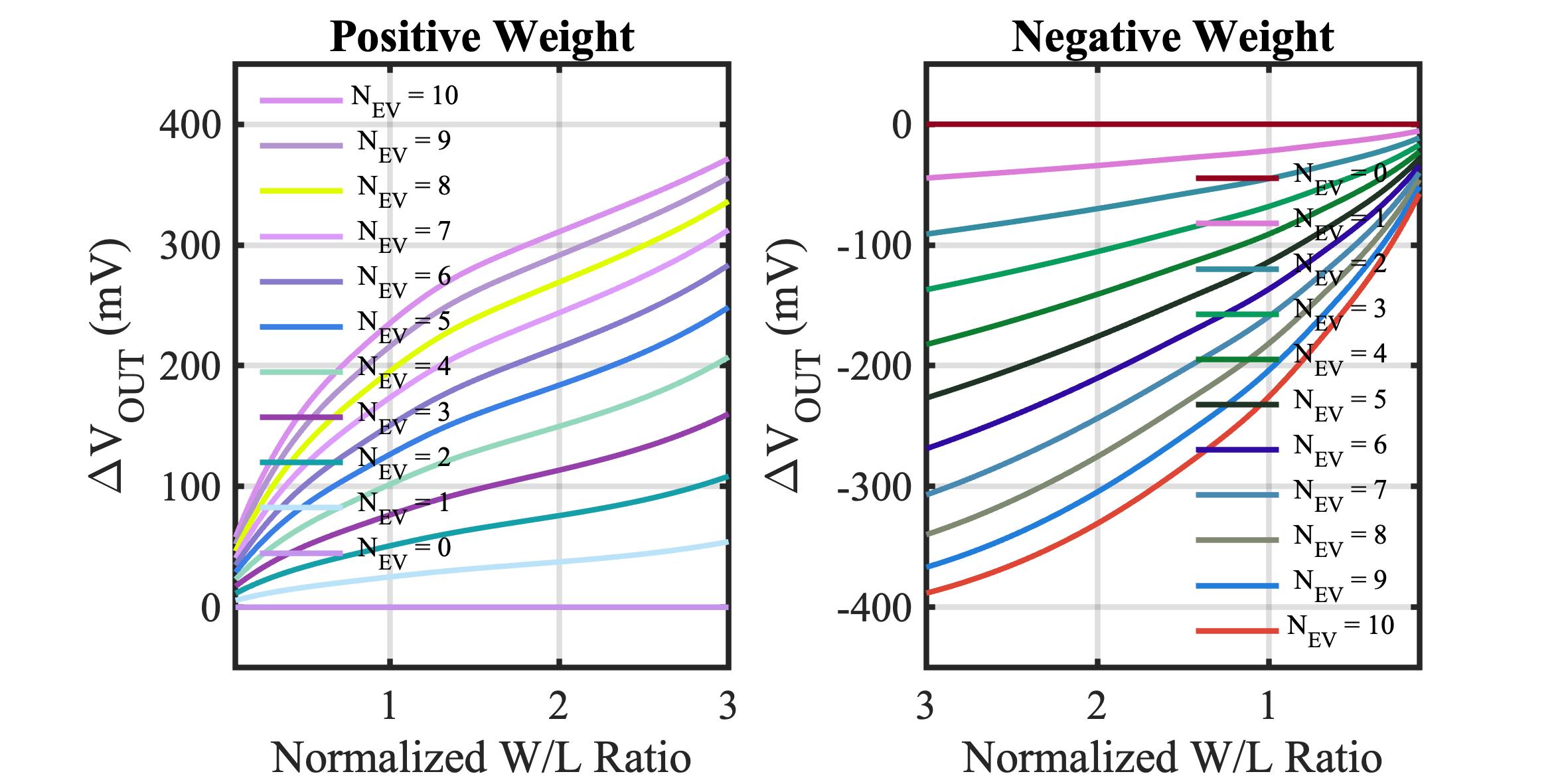}
\end{center}
\caption{Output accumulation voltage change (\si{\Delta V_{OUT}}) from the reset voltage of the kernel capacitor (\si{C_K}) as a function of normalized weight (normalized transistor \si{\frac{W}{L}} ratio) and input event spikes simulated on GF 22nm FD-SOI node for positive and negative weights.}
\label{vout_weight}
\end{figure}

To incorporate the circuit's nonideality in our algorithmic model, we have simulated the output characteristics of the positive and negative weights for the different numbers of input event spikes using the GF 22nm FD-SOI node. Figure \ref{vout_weight} represents the output voltage change on the accumulation capacitor (\si{\Delta V_{OUT}}) as a function of the normalized weight transistor's \si{\frac{W}{L}} ratios and different numbers of input event spikes. The figures show that the accumulated voltage can step up (for positive weights) and down (for negative weights) and the size of the step is dependent on the weight transistor's \si{\frac{W}{L}} ratio. However, the step size dependency is non-linear, and the non-linearity is larger when the weights are large and the pre-step voltage is close to the supply rails. This can be attributed to the fact that the weight transistors (\si{M_W}) enter the triode region when their drain-to-source voltage is low, causing the charging (discharging) current to drop compared to the typical saturation current. However, the number of input events is sparse for the DVS dataset, and having large weight values for all the weights in a kernel is highly unlikely for a neural network model. Hence, the non-linear characteristics of the weight transistors do not cause any significant accuracy issues in our algorithmic model. Besides, the circuit's asymmetry due to utilizing different types of transistors (pMOS for positive weights and nMOS for negative weights) is also captured and included in our algorithmic model.

\subsection{In-situ Multi-Pixel Multi-Channel Convolution Operation}

\begin{figure}[!b]
\begin{center}
\includegraphics[width=\linewidth]{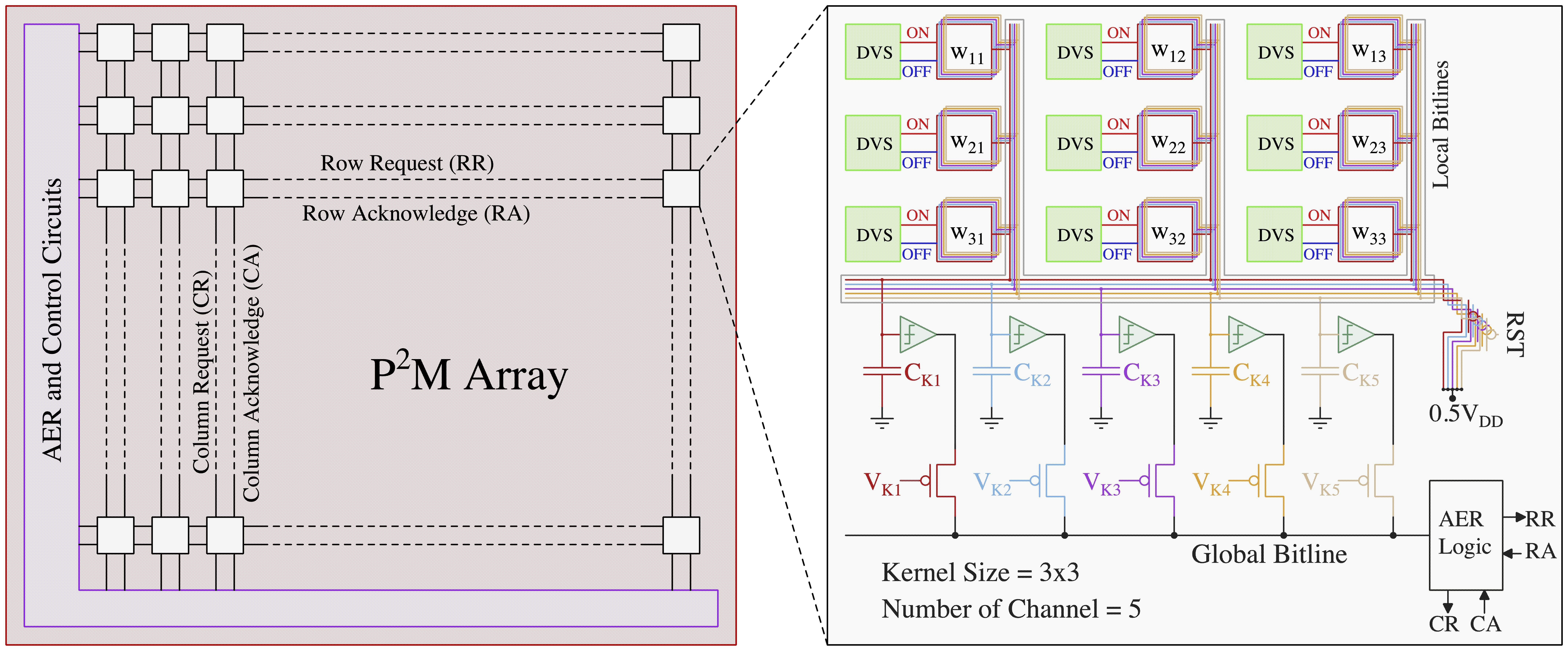}
\end{center}
\caption{Neuromorphic-P\si{^2}M array block diagram with peripheral control circuits and multi-channel configuration of our proposed neuromorphic-P\si{^2}M architecture.}
\label{p2m_arch}
\end{figure}

In the first spiking CNN layer, we have to perform \textit{spatio-temporal} MAC operations across multiple channels \textit{simultaneously} for each kernel. Figure \ref{p2m_arch} illustrates our proposed neuromorphic-P\si{^2}M architecture. The left sub-figure represents an array of DVS pixels (each white rectangular box includes multiple DVS pixels arranged in rows and columns) consisting of multiple channels distributed spatially. Each DVS pixel is connected with multiple weight transistors of the analog MAC blocks depending on the number of channels and stride (e.g., each DVS pixel will be connected with four sets of analog MAC blocks for a stride of 2). Each channel performs analog MAC operations asynchronously for a fixed temporal window (the length of each algorithmic time step). For instance, assume the kernel size is 3\si{\times}3 and each kernel has 5 different channels that are represented by the white rectangular boxes in the left sub-figure. The right sub-figure exhibits the zoomed version of the 3\si{\times}3 kernel with 5 different channels. Each channel has a dedicated accumulation capacitor (e.g., \si{C_{Ki}}, where i = 1, 2, ... 5) and a local bitline so that charge can accumulate across all the different channels at the same time. Depending on the kernel size, multiple weight transistors (both positive and negative) are connected to its kernel-dedicated accumulation capacitor using the local bitline of each channel. In this example, 18 weight transistors (kernel size = 3\si{\times}3 and for the ON and OFF channels of the DVS pixels) are connected with a single kernel capacitor. The per-channel accumulation capacitor and local bitline shared among the kernel's weight transistors enable simultaneously and massively parallel spatio-temporal MAC operations across different channels. The multiplication results (fixed amount of charge transfer to kernel capacitor from \si{V_{DD}} or from kernel capacitor to GND as a function of positive or negative weight depending on the weight values) accumulate on the kernel capacitor for a fixed temporal window (length of each algorithmic time step). These analog MAC operations are asynchronous and parallel across all the kernels for all the input feature maps (DVS pixels) throughout the sensor array. Finally, a thresholding circuit compares the final accumulated voltage on each channel's capacitor with a reference voltage to generate the output activation spike. Output activations from different channels are multiplexed (controlled by \si{V_{K1}}, \si{V_{K2}}, etc.) to communicate with the AER read circuits at the periphery (left sub-figure) through the kernel-level AER logic block (right sub-figure). The row request (RA) and row acknowledge (RA) signals are shared along the rows and the column request (CR) and column acknowledge (CA) signals are shared along the columns. 
After the read operation (described in subsection 2.3),  the kernel's accumulation capacitor is reset to 0.5\si{V_{DD}} by the reset transistor (\si{M_{RST}}) shown in Figure \ref{embedded_weight}. Note, the reset operation implies that there is no propagation of the voltage accumulated on the kernel's capacitor from one time step to subsequent time steps. Thus, the kernel capacitor voltage is unlike the typical representation of the membrane potential found in the literature \cite{datta_frontiers,datta2021training,spiking_lstm}, which is conserved across time steps. Taking into cognizance the above behavior, for the first layer of the network, we ensure our algorithmic framework includes thresholding and reset operation across time steps, thus faithfully representing the circuit behavior in algorithmic simulations.

\begin{figure}[!b]
\begin{center}
\includegraphics[width=0.8\linewidth]{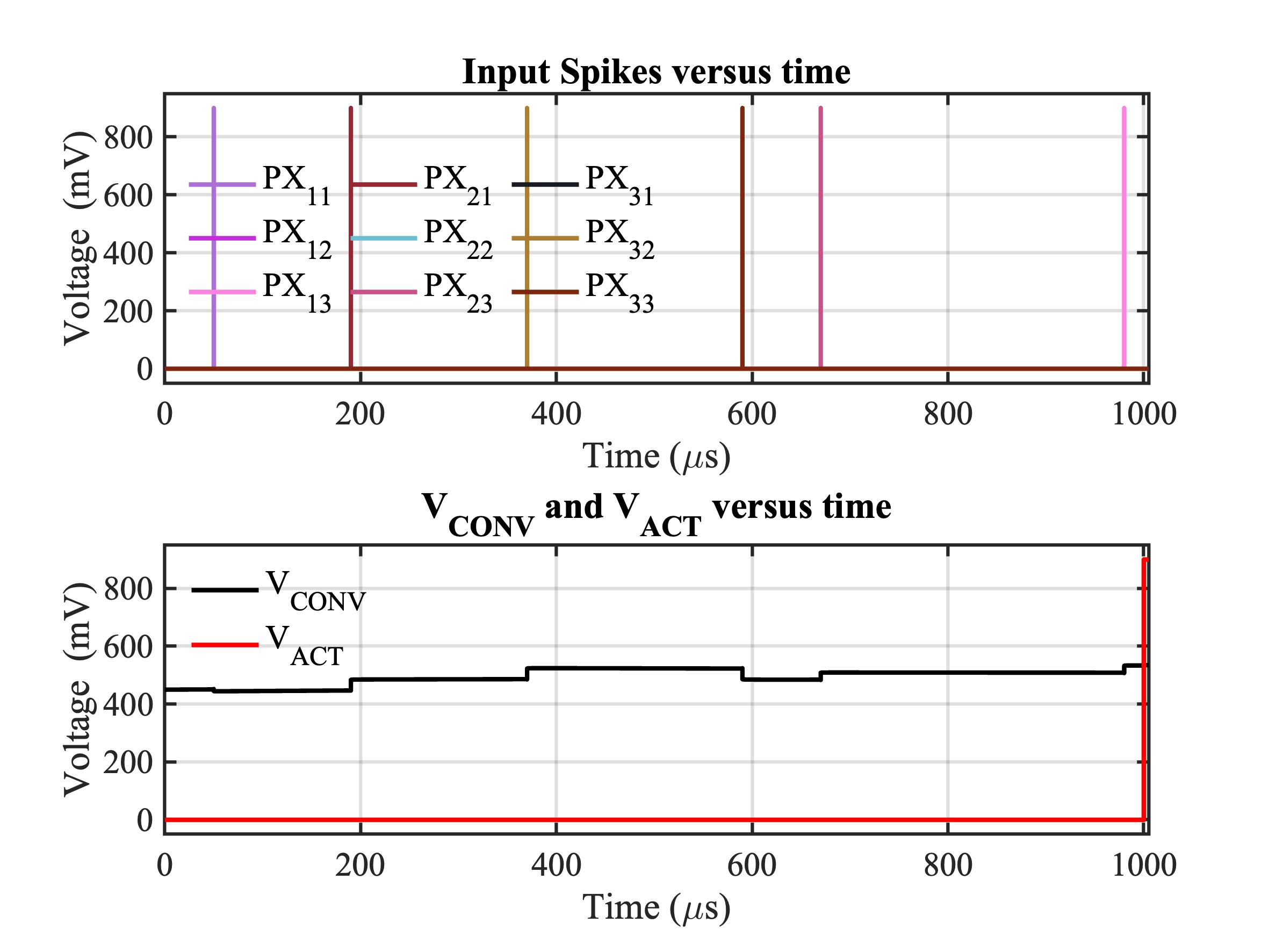}
\end{center}
\caption{A random convolution operation with output activation spike simulated on GF 22nm FD-SOI node.}
\label{conv_test}
\end{figure}

The frequency of the reset operation is based on the amount of time the capacitor can hold the charge without significant leakage. To minimize the capacitor leakage, we use high-\si{V_T} weight transistors, a switching transistor (\si{M_{SW}}) to disconnect the kernel's capacitor from the weight transistors, and kernel-dependent nullifying current source (\si{I_{NULL}}) (shown in Figure \ref{embedded_weight}). According to our HSpice simulations, in the worst-case scenario (all weights are maximum in the kernel, which is very unlikely in the neural network model), the voltage on the accumulation capacitor deviates due to leakage from its ideal value by a mere 22 mV over a significantly longer duration of time (e.g., 1 ms). Based on the reset frequency, the length of each algorithmic time step of our neural network model has been set to 1 ms for the first layer.

Figure \ref{conv_test} illustrates an asynchronous convolution and output activation spike generation example of our proposed neuromorphic-P\si{^2}M using the GF22nm FD-SOI technology node considering random inputs and weights. For this simulation, a kernel size of 3\si{\times}3 has been considered. The weights, the instant of the events, and the number of events per DVS pixel are generated randomly. 
For the test HSpice simulation, 1 ms simulation time has been considered according to our algorithmic framework, hence, all the output events from DVS pixels within this time period will be multiplied with their weights and accumulated on the Kernel's accumulation capacitor before being compared with a fixed threshold voltage. 
The top subplot exhibits that the DVS pixels (e.g., PX\si{_{11}}, PX\si{_{21}}, etc.) are generating the event spikes at different time instants. PX\si{_{13}}, PX\si{_{21}}, PX\si{_{23}}, PX\si{_{32}} are connected with positive weights, whereas, the other pixels are connected with negative weights. It may also be noted, a few pixels (e.g., PX\si{_{12}}, PX\si{_{22}}, PX\si{_{31}}) do not generate any event during this time frame. These no-event generation scenarios are also considered in this test simulation to mimic the actual dataset sparsity. From the bottom subplot, it can be observed that the convolution output (\si{V_{CONV}}) of our analog MAC circuit is updating (charging or discharging) for each input event spike. When the weight is positive (negative), the accumulation voltage steps up (down) depending on the weight value. Finally, after the fixed temporal window, the convolution output has been compared with the threshold voltage. If the convolution output is higher than the threshold voltage, the comparator will generate an output activation spike (\si{V_{ACT}}) for the next layer for each kernel. 

\subsection{P\si{^2}M Address-Event Representation (AER) Read Operation}
\begin{figure}[!b]
\begin{center}
\includegraphics[width=1\linewidth]{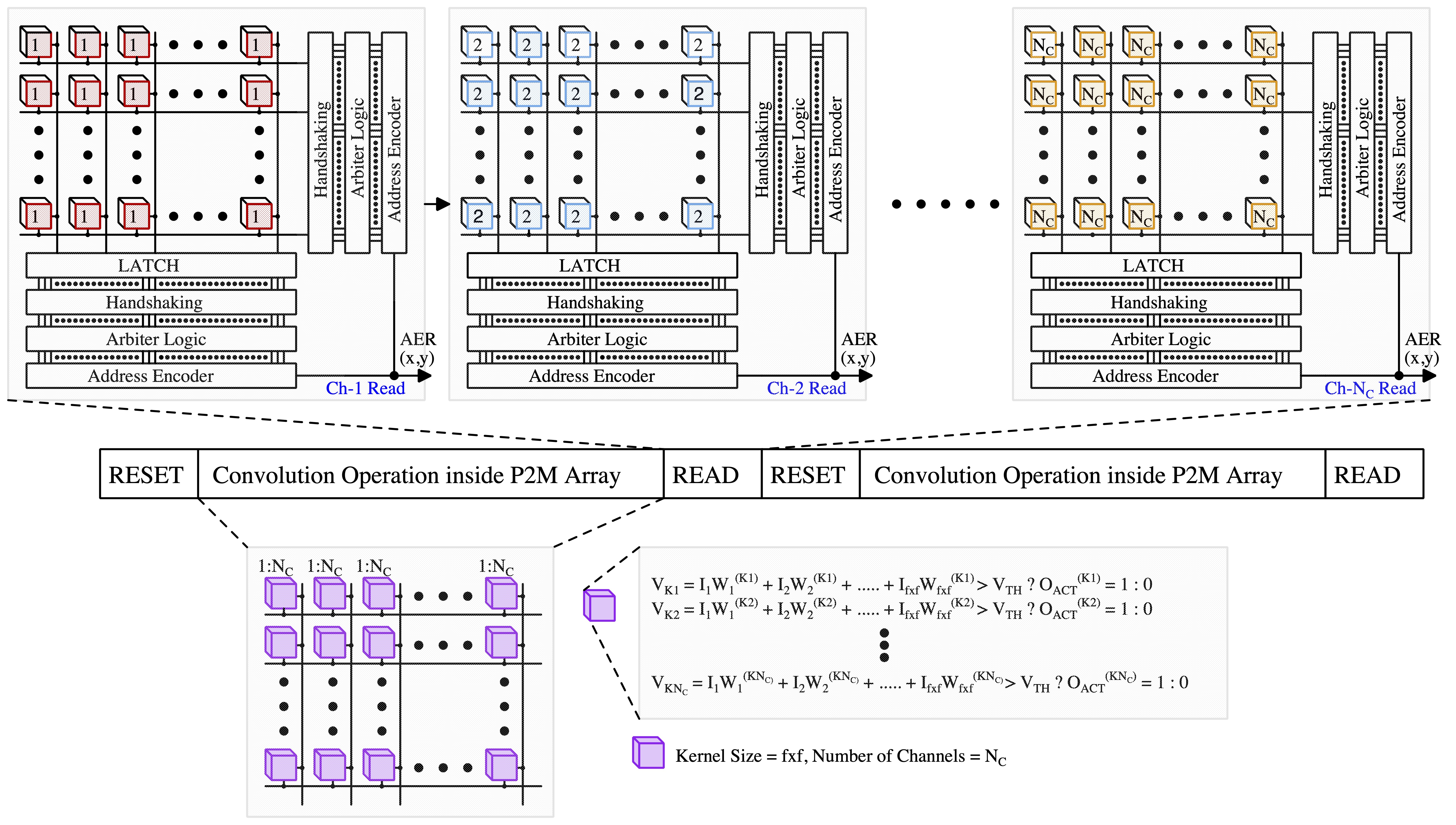}
\end{center}
\caption{AER read scheme of our proposed neuromorphic-P\si{^2}M architecture.}
\label{aer_read}
\end{figure}

In this sub-section, we propose modifications to the standard AER scheme in a manner so that it can be compatible with the presented asynchronous processing in-pixel computations.
We are utilizing the asynchronous AER read-out scheme \cite{AER1} to read the output activations from the first convolution layer (mapped onto the DVS pixels using our proposed neuromorphic-P\si{^2}M paradigm). The representative read scheme is illustrated in Figure \ref{aer_read}. Our P\si{^2}M architecture can support multiple numbers of channels (e.g., \si{N_C}) as required by the spiking CNN model. The outputs of the channels (thresholded output activation spikes) are read sequentially throughout the P\si{^2}M array in an asynchronous manner. At a time, one channel is being asserted of the P\si{^2}M array by activating \si{V_{Ki}} sequentially, where, i = 1, 2, ... \si{N_C} (shown in Figure \ref{p2m_arch}). Kernel-level AER logic block shared among different channels for each spatial feature map generates the row and column request signals whenever there is an output activation spike in the kernel. For AER reading, row-parallel techniques can be used where it latches all the events generated in a single row and read them sequentially \cite{AER1}. The peripheral address encoders (row and column encoders) of the AER read circuits output the x and y address of the output activation in parallel. Moreover, while performing the read operation, we can also pipeline the next reset and convolution phases without waiting for the read phase to be completed by adding a transistor between the kernel capacitor and the comparator. The comparator output can be stored on the dynamic node for a short period of time or even we can use a small holding capacitor to hold the output activation for a sufficient amount of time considering the read operation. As the output activations are sparse and AER read can be completed within a few \si{\mu s} windows, we can also utilize our architecture to perform the convolution and read phase in parallel. Besides, due to performing the in-pixel convolution operation, the output activation map size is reduced as a function of the kernel size and number of strides. In addition, we also do not need to send an extra bit to define the polarity of the event (ON or OFF-event) similar to the base DVS systems. As a result, the required number of address bits that need to be communicated off-chip has been reduced from the base DVS system. Hence, our P\si{^2}M architecture maintains the energy benefit of a sparse system due to utilizing the AER read scheme along with lower off-chip communication energy cost due to generating fewer address bits per output activation. 

\subsection{Process Integration and Area Consideration}

\begin{figure}[!t]
\begin{center}
\includegraphics[width=0.4\linewidth]{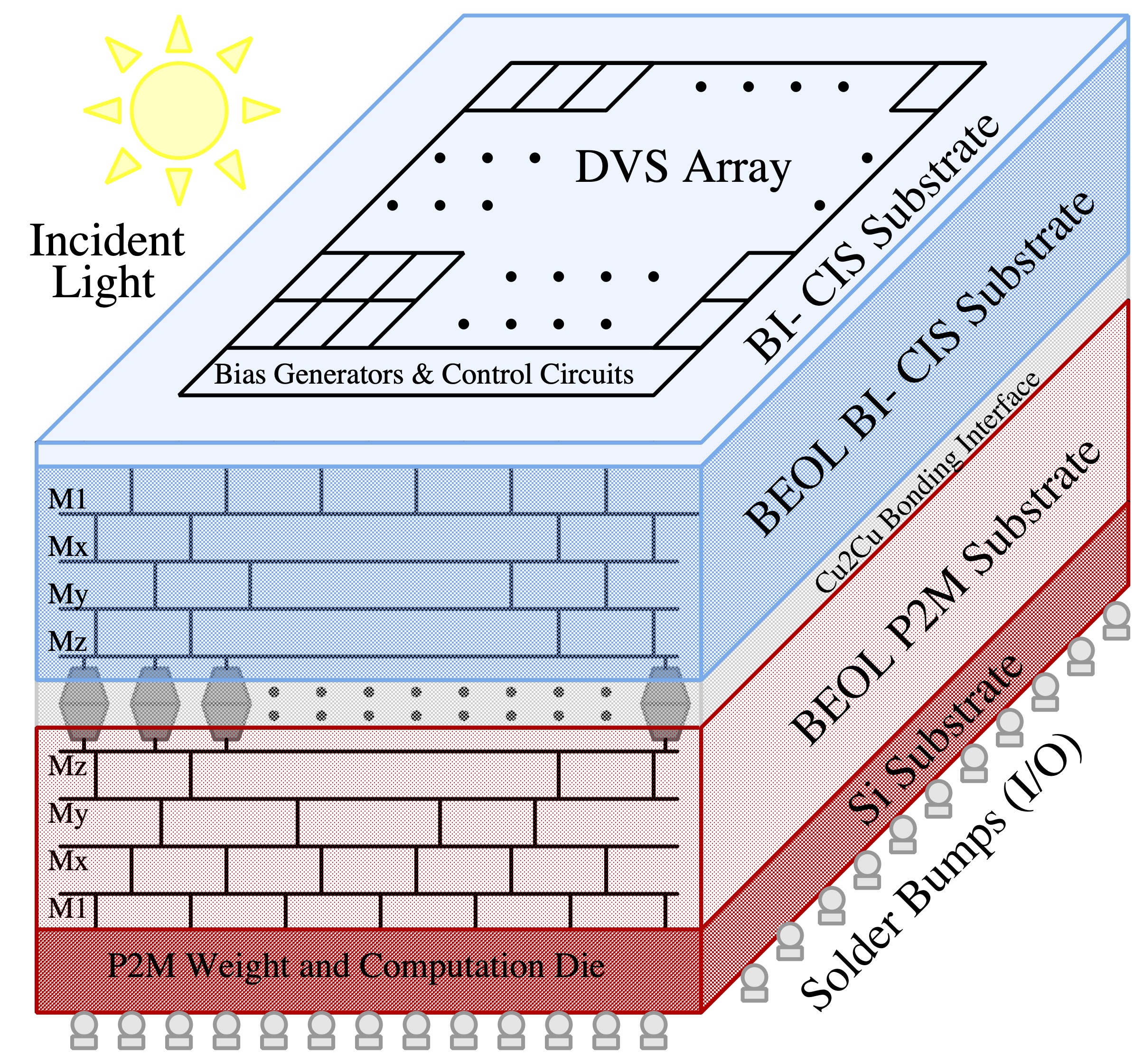}
\end{center}
\caption{Representative illustration of a heterogeneously integrated system featuring neuromorphic-P\si{^2}M paradigm utilizing Cu2Cu Bonding.} 
\label{p2m_3D}
\end{figure}

Figure \ref{p2m_3D} exhibits the representative illustration of a heterogeneously integrated system featuring our proposed neuromorphic-P\si{^2}M paradigm. Our proposed system can be divided into two key dies, i) a backside illuminated CMOS image sensor (BI-CIS), consisting of the DVS pixels and biasing circuitry, and ii) a die containing multi-bit multi-channel weight transistors, accumulation capacitors, comparators, and AER read circuits. From Figure \ref{p2m_arch}, it can be observed that for each spatial feature (DVS pixels), the algorithm requires multiple channels that incur higher area due to multiple weight transistors and one accumulation capacitor per channel. However, due to the advantages of heterogeneous integration, our bottom die can be fabricated on an advanced technology node compared to the top die (BI-CIS). Hence, multiple channels in the bottom die can be accommodated and aligned with the top die without any area overhead while maintaining the neural network model accuracy. It may be noted that typical DVS pixels are larger due to the inclusion of a capacitive feedback difference amplifier. The overall system can be fabricated by a wafer-to-wafer bonding process using pixel-level hybrid Cu2Cu interconnects \cite{C2C_ref2}, \cite{C2C_ref3}, \cite{C2C_ref1}. Each DVS pixel has two Cu2Cu interconnects for its ON and OFF-channel, respectively. Considering the DVS pixel area of 40 \si{\mu m}\si{\times}40 \si{\mu m} \cite{DVS_ref1} for 128\si{\times}128 sensor array, Cu2Cu hybrid bonding pitch of 1 \si{\mu m} \cite{C2C_pitch} and the analog convolution elements (weight transistors, comparators, accumulation capacitors) area in GF22nm FD-SOI node, our neuromorphic-P\si{^2}M architecture can support maximum of 128 and 32 channels with a kernel size of 3\si{\times}3 for stride 2 and 1, respectively. However, in our algorithmic framework, 32 channels with stride 2 have been utilized. 
Such kernel-parallel MAC structure allows us to enable in-situ convolution operation without a need for weight transfer from a different physical location, thus this method does not lead to any data bandwidth or energy bottleneck.

\section{P\si{^2}M-constrained Algorithm-Hardware Co-design}

\begin{figure}[!t]
\begin{center}
\includegraphics[width=0.7\linewidth]{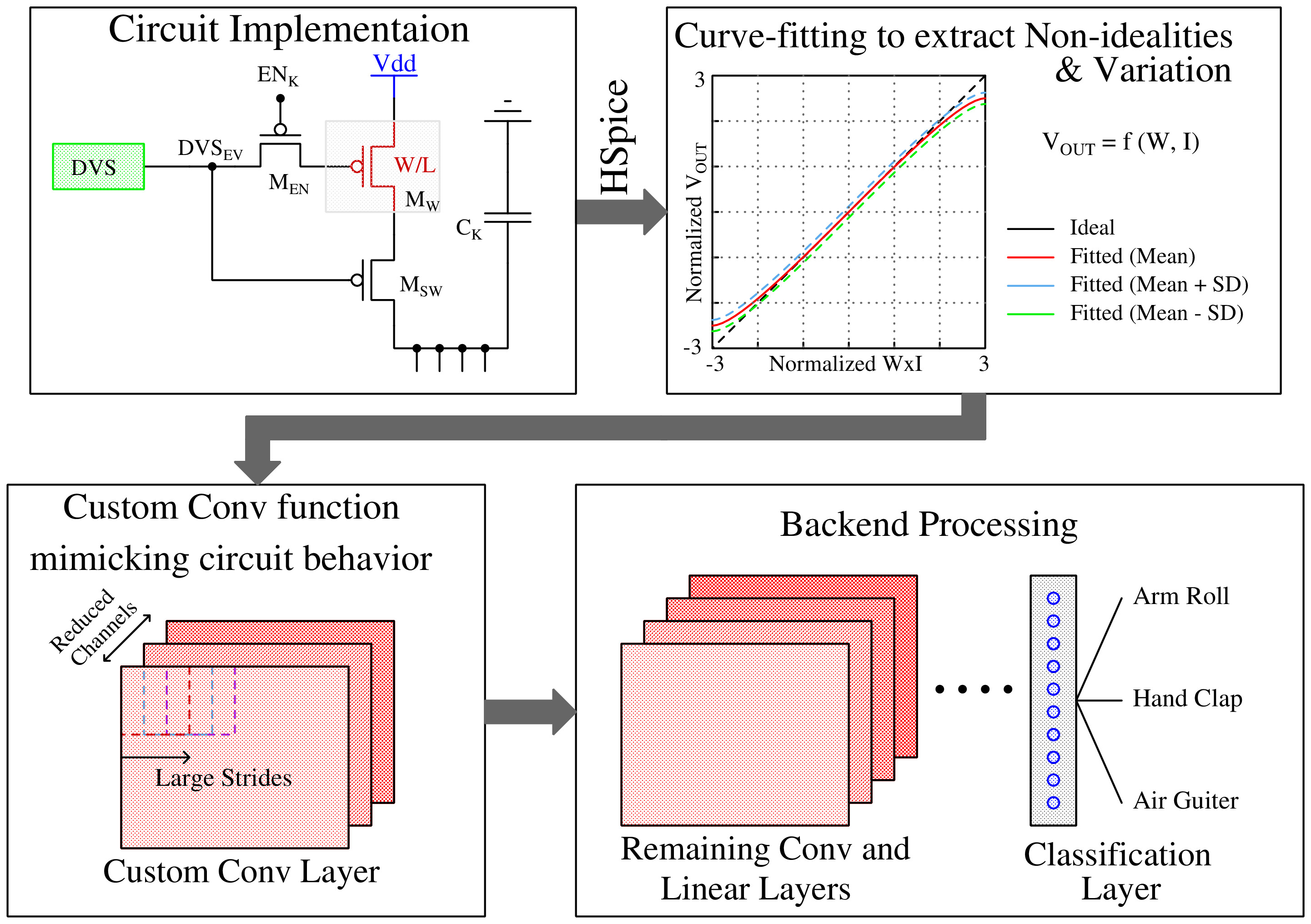}
\end{center}
\caption{Hardware-Algorithm co-design framework to enable our proposed neuromorphic-P\si{^2}M approach.}
\label{codesign_framework}
\end{figure}

In this section, we present our algorithmic framework implementation guided by our proposed neuromorphic-P\si{^2}M architecture. The in-pixel charge-based analog convolution generates non-ideal non-linear convolution in addition process variation yields a deviation of the convolution result from the ideal output. Moreover, leakage poses constraints on the maximum length of each algorithmic time step, and the area limits the number of channels utilized per each spatial feature map. The hardware-algorithm co-design framework of our proposed neuromorphic-P\si{^2}M approach has been illustrated in Figure \ref{codesign_framework}. More details on including non-idealities, process variation, leakage, and area effects in our algorithmic framework are given in the following subsections. 

\subsection{Custom Convolution for the First Layer Modeling Circuit Non-linearity and Process Variation}

\begin{figure}[!b]
\begin{center}
\includegraphics[width=0.8\linewidth]{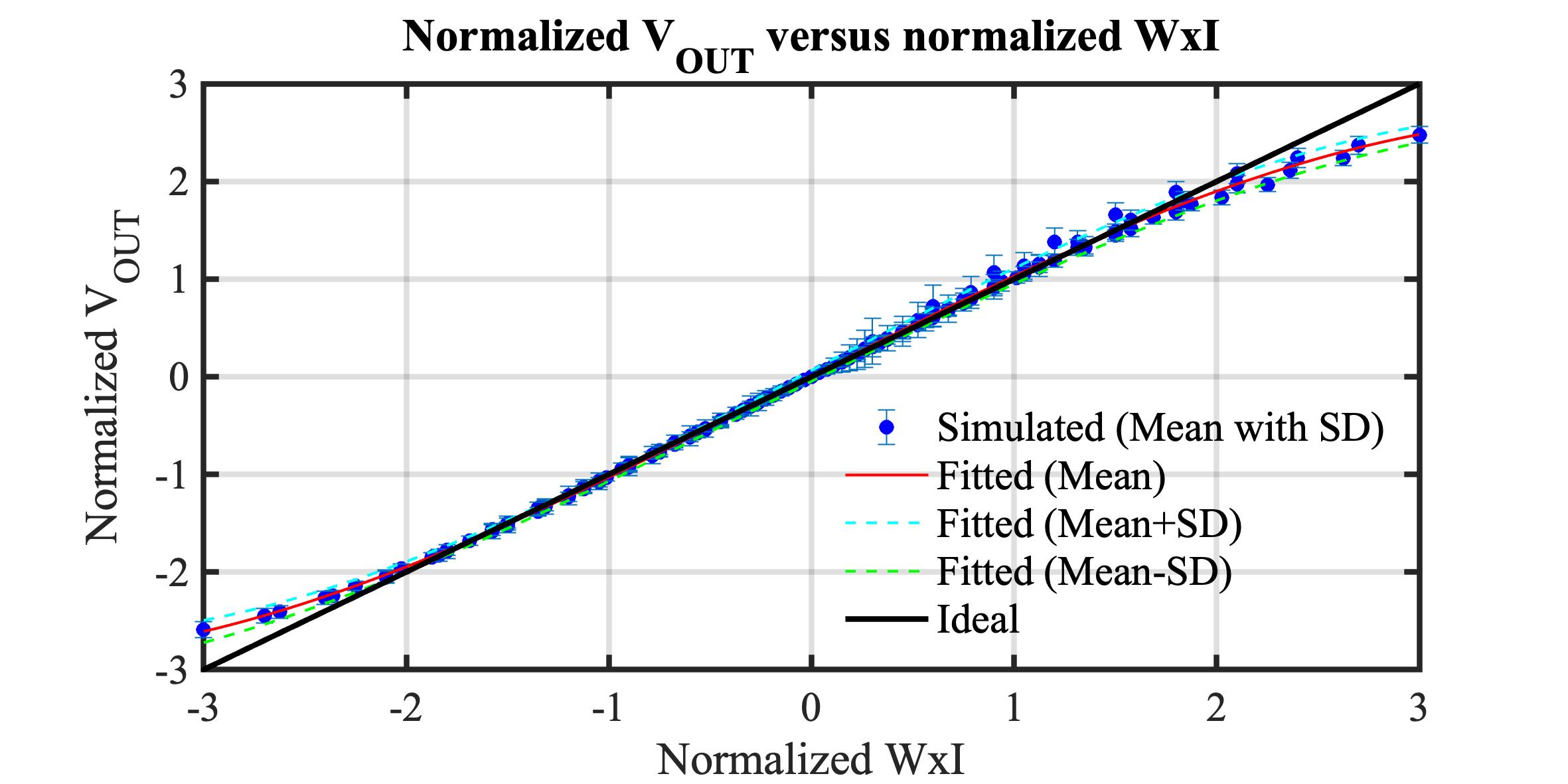}
\end{center}
\caption{A scatter plot with standard deviation comparing the pixel output voltage to ideal multiplication value of Weights\si{\times}Input activation (Normalized W\si{\times}I).} 
\label{scatter_plot}
\end{figure}

From an algorithmic perspective, the first layer of a spiking CNN is a linear convolution layer followed by a non-linear activation unit. In our neuromorphic-P\si{^2}M paradigm, we have implemented the weights utilizing voltage accumulation through appropriately sized transistors that are inherently non-linear. As a result, any analog convolution circuit built on transistor devices will exhibit non-ideal non-linear behavior. Hence, to suppress the non-linearity, we have tuned our weights (transistor's geometry) in a non-linear manner in such a way that the output accumulation voltage steps can increase or decrease linearly for positive and negative weights, respectively. However, the nonlinearity is also a function of the drain-to-source voltage of the weight transistors. In our scheme, we are charging or discharging the kernel's capacitor during the computation phase. The charging and discharging current depending on the weight values are functionally dependent on the drain-to-source voltage. Hence, when the accumulation voltage on the (\si{V_{OUT}} node in Figure \ref{embedded_weight}) gets larger (smaller) for the positive (negative) weights, the transistor enters into the triode region, hence, the charging or discharging current reduces. Besides, the exact same positive and negative weight values cannot ensure the exact same change in voltage accumulation due to device asymmetry (pMOS for positive weight implementation and nMOS for negative weight implementation). Furthermore, due to process variation, the transistor's geometry cannot be fabricated precisely, hence, the convolution output current can also vary due to process variation. Considering all these non-linear non-ideal behaviors and process variations, we extensively simulated our proposed P\si{^2}M paradigm for a wide range of input spikes and weights combinations considering leakage and around 3-sigma variation using GF22nm FD-SOI technology node. Figure \ref{scatter_plot} illustrates the resulting HSpice results with a standard deviation bar, i.e. the normalized convolution output voltages per pixel corresponding to a range of weights and input number of spikes, which have been modeled using a behavioral curve-fitting function. 
Note, for the scatter plot we have used 100 \si{\mu s} temporal window for the convolution phase to save the total circuit simulation time as we have to run 1000 Monte-Carlo simulations for each combination of weights and the number of input spikes.
In our algorithmic framework, a random gaussian sample value has been generated between the mean\textpm  standard deviation for each particular normalized weight times input event value to capture the effects of the process variation. For the fixed simulation time for the event stream, in each Kernel, the accumulation output voltage per pixel is calculated first, then added to the other pixel's accumulation voltage inside the kernel to calculate the final output. The algorithmic framework was then used to optimize the spiking CNN training for the event-driven neuromorphic datasets.  

\begin{figure}[!t]
\begin{center}
\includegraphics[width=0.9\linewidth]{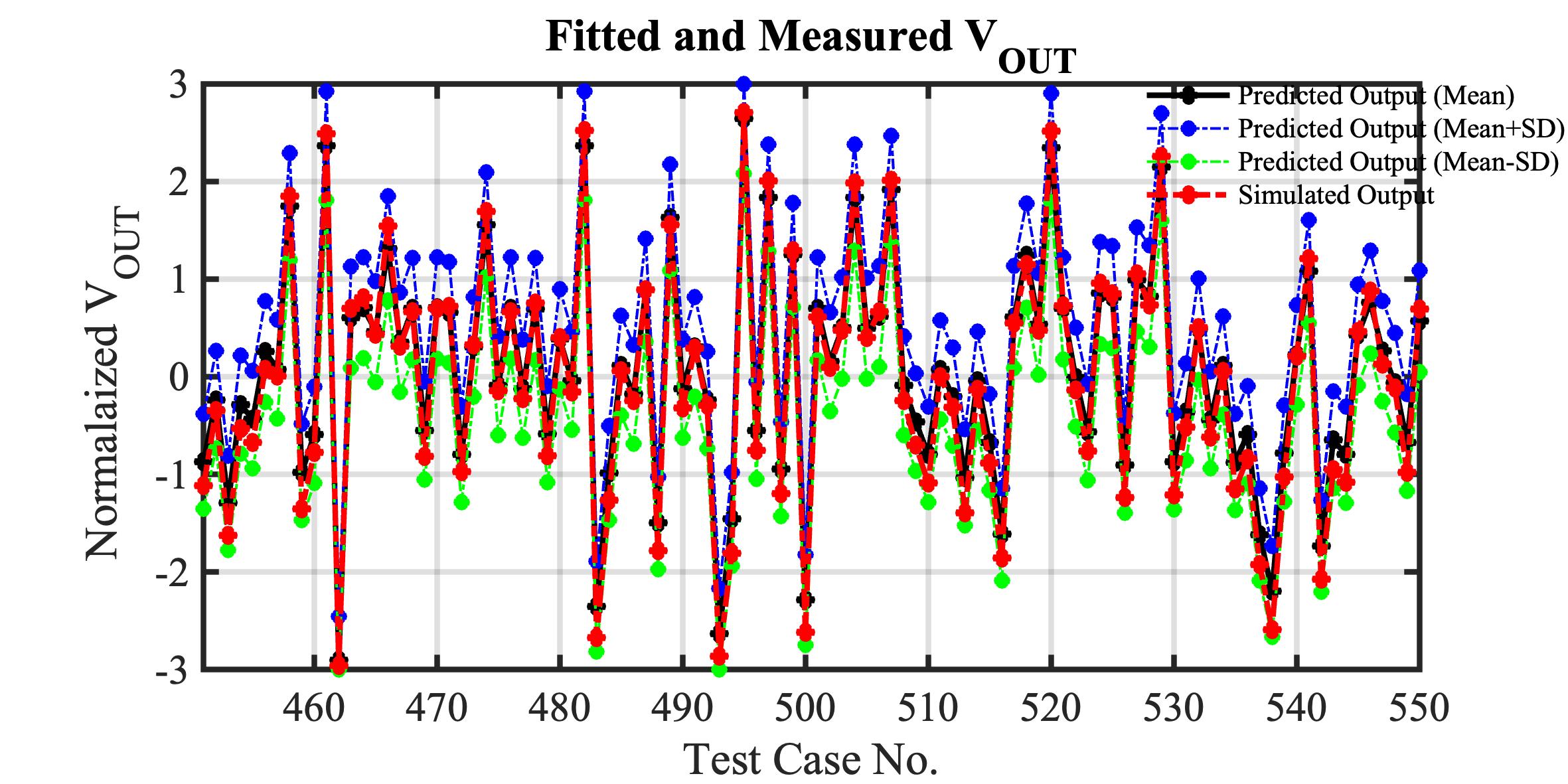}
\end{center}
\caption{100 random HSpice simulation results for 3x3 Kernel benchmarking with the fitted equations.}
\label{random_tests}
\end{figure}

To validate our Hspice simulations generated curve-fitting function's prediction accuracy, we have tested 1000 random cases. In these test cases, we have used a kernel size of 3\si{\times}3, where the weight values are generated randomly. Moreover, the number of input event spikes and time instants for the input spikes are also randomly generated. 
Note, these random tests utilize 100 \si{\mu s} of simulation time for the convolution phase to reduce the total simulation time. As we have earlier mentioned, utilizing kernel-dependent nullifying current source, high-\si{V_T} weight transistors, and a switch to disconnect the kernel's capacitor exhibits a maximum of 22 mV error in the worst-case scenario. Hence, random HSpice tests ignoring a long time (1 ms of simulation time length for each event stream of our neural network model) leakage will not incur any significant accuracy issues for these 1000 random tests. 
Among 1000 random tests, only 100 test results (for clear visibility) are shown in Figure \ref{random_tests}. In the figure, the curve-fitted mean and mean\textpm standard deviation predictions of our proposed analog MAC operations are shown with HSpice-generated simulation results.  We have used a 3rd order single variable (normalized weight times input event spikes) polynomial to generate the curve fitting functions (mean, mean\textpm standard deviation) considering 0.55\% mean RMSE of our analog MAC to minimize the computation complexity in our algorithmic framework while maintaining high accuracy. It can clearly be seen that the predicted mean output follows the Hspice results closely, and the HSpice outputs fall between the mean\textpm standard deviation value.

\subsection{Circuit-Algorithm Co-optimization of spiking CNN Backbone subject to P\si{^2}M Constraints}

In our proposed neuromorphic-P\si{^2}M architecture, we have utilized a kernel-dedicated capacitor to enable instantaneous and massively parallel spatio-temporal convolution operation across different channels. We need a kernel-dedicated capacitor to preserve the temporal information of input DVS spikes across different channels simultaneously. Moreover, there is a direct trade-off between the acceptable leakage and capacitance value (a large capacitor incurs a large area, however, results in lower leakage). Almost 47\% of the area in our  P\si{^2}M array is occupied by the capacitors.  Hence, we have reduced the number of channels in our spiking CNN models compared to the baseline neural architecture not to incur any area overhead while preserving the model accuracy. In addition, the leakage also limits the length of each algorithmic time step in our algorithmic framework. We have also reduced the time length in our neural network model to minimize the kernel-dependent leakage error of our custom first convolution layer. Moreover, to reduce the amount of data transfer between the P\si{^2}M architecture and the backend hardware processing of the remaining spiking CNN layers, we have avoided the max pooling layer and instead used a stride of 2 in the P\si{^2}M convolutional layer. Lastly, we incorporate the Monte Carlo variations in the proposed non-linear custom convolutional layer explained above in our algorithmic framework. In particular, we have estimated the mean and standard deviation of the output of the custom convolutional layer from extensive circuit simulations. We then train our spiking CNN with the addition of the standard deviation as noise to the mean output of the convolutional layer. This noise addition during training is crucial to increase the robustness of our spiking CNN models, as otherwise, our models would incur a drastic drop in test accuracy.

\section{Experimental Results}

\subsection{Benchmarking Dataset and Model}

This article focuses on the potential use of P\si{^2}M for event-driven neuromorphic tasks where the goal is to classify each video sample captured by the DVS cameras. In particular, we evaluate our P\si{^2}M approach on two large-scale popular neuromorphic benchmarking datasets. 

\textbf{DVS128-Gesture}: The IBM DVS128-Gesture \cite{dvs-gesture} is a neuromorphic gesture recognition dataset with a temporal resolution in \si{\mu s} range and a spatial resolution of \si{128{\times}128}. It consists of 11 gestures (1000 samples each), such as hand clap, arm roll, etc., recorded from 29 individuals under three illumination conditions, and each gesture has an average duration of 6 seconds. To the best of our knowledge, it is the most challenging open-source neuromorphic dataset with the most precise temporal information. 

\textbf{NMNIST}: The neuromorphic MNIST \cite{nmnist} dataset is a converted dataset from MNIST. It consists of 50K training images and 10K validation images. We preprocess it in the same way as in N-Caltech 101. We resize all our images to \si{34{\times}34}.

For these datasets, we apply a 9:1 train-valid split. We use the Spikingjelly package \cite{SpikingJelly} to process the data and integrate them into a fixed time interval of 1 ms based on the kernel's capacitor retention time supported by our neuromorphic-P\si{^2}M circuit. However, such a small integration time would lead to a large number of time steps for the neuromorphic datasets considered in this work whose input samples are at least a few seconds long. This would significantly exacerbate the training complexity. To mitigate this concern, we first pre-train a spiking CNN model with a large integration time in the order of seconds (i.e., with a small number of time steps) without any P\si{^2}M circuit constraints. We then decrease the integration time of the first spiking convolutional layer for P\si{^2}M implementation and integrate the spikes in the second interval such that the network from the second layer processes the input with only a few time steps. We fine-tune this network from the second layer while freezing the first layer since training the first layer significantly increases the memory complexity due to a large number of time steps. This is because the gradients of the first layer need to be unrolled across all the time steps. We use four convolutional layers, followed by two linear layers at the end with \si{512} and \si{10} neurons respectively. Each convolutional layer is followed by a batch normalization layer, spiking LIF layer, and max pooling layer.

\begin{table}[!b]
\caption{Comparison of the test accuracy of our P\si{^2}M enabled spiking CNN models with the baseline spiking CNN counterparts, where `MP' denotes membrane potential, `Custom Conv.' denotes the incorporation of the non-ideal model to the ML algorithmic framework, and `Reduced dimensionality' denotes the reduction in the number of channels in the first convolutional layer.}
\label{tab:dvs_accuracy_results}
\begin{center}
\setlength{\tabcolsep}{1.5mm}{
\begin{tabular}{l|c|c|c|c}
\hline
\hline
Dataset & MP 1$^{st}$ layer & Custom Conv. & Reduced dimensionality & Accuracy ($\%$) \\
\hline
DVS128-Gesture & \centered{\cmark} & \centered{\xmark}  & \centered{\xmark} &  93.40 \\
\hline
DVS128-Gesture & \centered{\xmark} & \centered{\xmark}  & \centered{\xmark} &  88.78 \\
\hline
DVS128-Gesture & \centered{\xmark} & \centered{\cmark}  & \centered{\xmark} &  88.54 \\
\hline
DVS128-Gesture & \centered{\xmark} & \centered{\cmark}  & \centered{\cmark} &  88.36 \\
\hline
NMNIST & \centered{\cmark} & \centered{\xmark}  & \centered{\xmark} &  98.10 \\
\hline
NMNIST & \centered{\xmark} & \centered{\xmark}  & \centered{\xmark} &  93.68 \\
\hline
NMNIST & \centered{\xmark} & \centered{\cmark}  & \centered{\xmark} &  93.44 \\
\hline
NMNIST & \centered{\xmark} & \centered{\cmark}  & \centered{\cmark} &  93.12 \\
\hline
\end{tabular}
}
\end{center}
\end{table}

\subsection{Classification Accuracy}

We evaluated the performance of the baseline and P\si{^2}M custom spiking CNN models on the two datasets illustrated above in Table \ref{tab:dvs_accuracy_results}. Note that all these models are trained from scratch. As we can see, the custom convolution model does not incur any significant drop in accuracy for any of the two datasets. However, the removal of the state variable i.e., the membrane potential in the first layer leads to $\sim 5\%$ drop in test accuracy on average. This might be because of the loss in the temporal information of the input spike integration from the DVS camera. Additional P\si{^2}M constraints such as less number of channels and increased strides in the first convolutional layer (see Section 3.2) hardly incur any additional drop in accuracy. Overall, our P\si{^2}M-constrained models lead to a $5.2\%$ drop in test accuracy on average across the two datasets.

\begin{table}[!b]
\caption{Energy estimation for different hardware components. The energy values are measured for designs in 22 nm CMOS technology. Note, the sensing energy (${E_{sens}}$) of our model includes the convolution energy for P\si{^2}M as the convolution is performed as a part of the sensing operation. The communication energy ($e_{comm}$) includes both the energy consumption of sending the address bits from the sensor to the transmitter ($e_{sens-to-tx}$) and wireless transmitter energy ($e_{tx}$). For ${e_{mac}}$ and ${e_{ac}}$, we convert the corresponding value in 45 nm to that of 22 nm by following standard scaling strategy \cite{stillmaker2017scaling}.}
\begin{center}
\scalebox{0.9}{
\begin{tabular}{|c|c|c|c|c|}
\hline
Model type & Sensing Energy (mJ) & Comm Energy (pJ/bit) & MAC Energy (pJ) & MAdds Energy (pJ)  \\
{} & ($E_{sens}$)  & ($e_{comm} = e_{sens-to-tx} + e_{tx} $) & ($e_{mac}$) & ($e_{ac}$) \\ 
\hline
\hline
P$^2$M (ours) & 26.588 & 4.1 & 1.568  & 0.03  \\
  \hline
  Baseline & 26.032 & 4.1 & 1.568 & 0.03 \\
 \hline
\end{tabular}}
\end{center}
\label{tab:energy_estimates}
\end{table}

\begin{figure}[!b]
\begin{center}
\includegraphics[width=\linewidth]{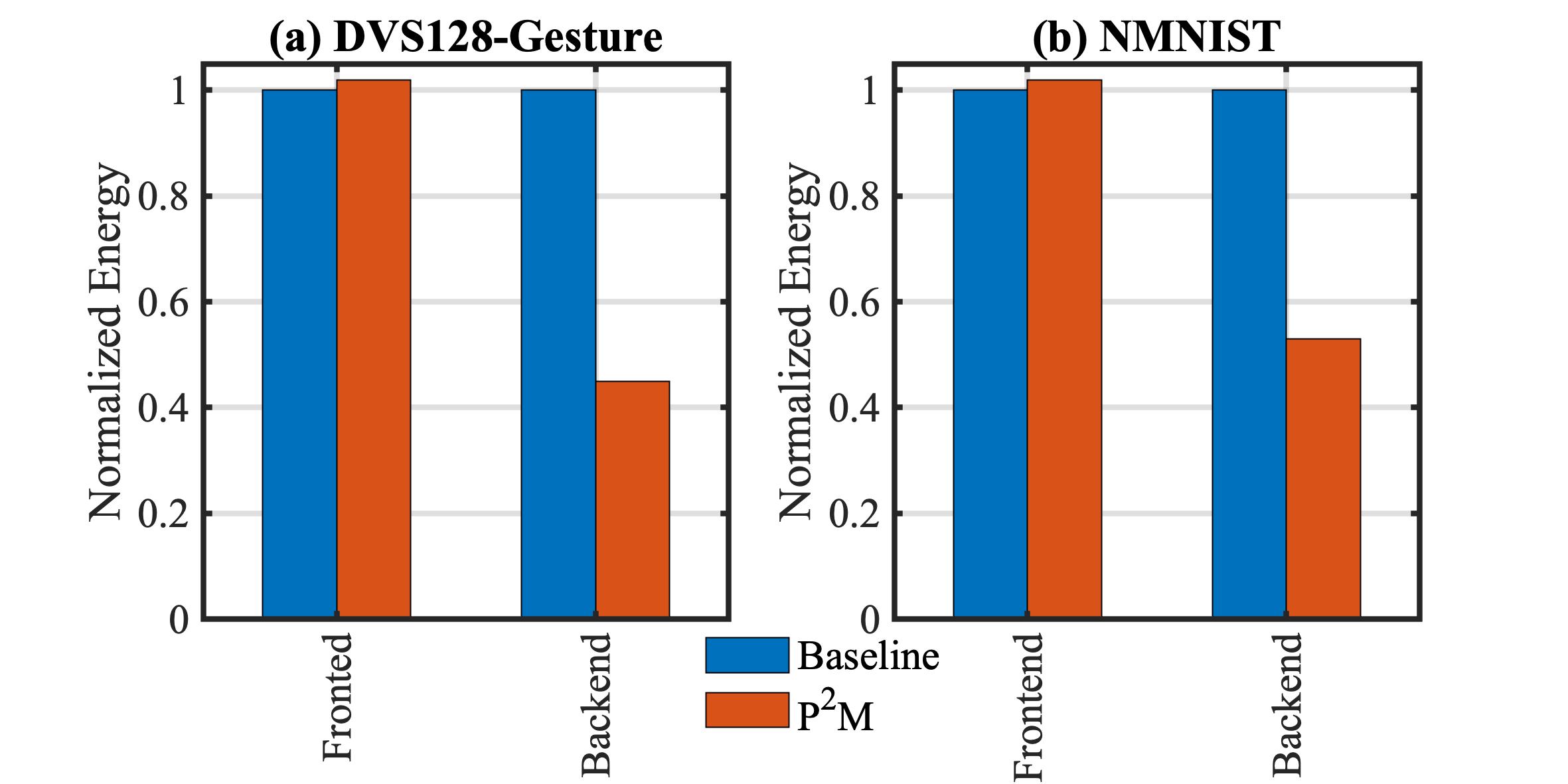}
\end{center}
\caption{Comparison of the energy consumption between baseline and P\si{^2}M implementations of spiking CNNs to process neuromorphic images from (a) DVS128-Gesture, and (b) NMNIST datasets.}
\label{fig:energy_comparison}
\end{figure}

\subsection{Analysis of Energy Consumption}

We develop a circuit-algorithm co-simulation framework to characterize the energy consumption of our baseline and P\si{^2}M-implemented spiking CNN models for neuromorphic datasets. Note that we do not evaluate the latency of our models since that would depend heavily on the underlying hardware architecture and data flow of the backend hardware (\textit{i.e.,} the hardware processing the remaining layers of the CNN, excluding the first layer that is processed using our P\si{^2}M paradigm). 
The frontend energy ($E_{frontend}$) is comprised of sensor energy ($E_{sens}$), and communication energy ($E_{com}$), while the backend energy ($E_{backend}$) to process the SNN layers (excluding the first layer for the P\si{^2}M implementation) is primarily composed of the accumulation operations incurred by the spiking convolutional layers ($E_{ac}$) and the parameter read ($E_{read}$) costs. Assuming $T$ denotes the total number of time steps and $s$ denotes the sparsity. The energy components can be approximated as:

\begin{align}
E_{frontend} & \approx \underbrace{e_{event}*N_{event}+E_{bias}}_{E_{sens}} + 
\underbrace{(e_{sens-to-tx}+e_{tx})*N_{event}}_{E_{com}} \\
E_{backend} & \approx \underbrace{e_{ac}*N_{ac}*s*T}_{E_{ac}} + \underbrace{e_{read}*N_{read}}_{E_{read}}
\end{align}

Here, $e_{event}$ represents per-pixel sensing energy, $N_{event}$ denotes the number of events communicated from the sensor to the backend, and $E_{bias}$ is the biasing energy for the DVS pixel array considering the dataset duration. In addition, $e_{sens-to-tx}$ is the communication energy to send the address bits from the sensor node to the transmitter, and $e_{tx}$ is the wireless transmission energy to the backend. Note that the first convolutional layer of the SNN in the baseline implementation requires MAC operations, and hence, we need to replace $e_{ac}$ with the MAC energy $e_{mac}$ and use $s{=}1$. For a spiking convolutional layer that takes an input $\mathbf{I} \in R^{h_i{\times}w_i{\times}c_i}$ and weight tensor $\mathbf{\theta} \in R^{k{\times}k{\times}c_i{\times}c_o}$ to produce output $\mathbf{O} \in R^{h_o{\times}w_o{\times}c_o}$, the $N_{ac}$ \cite{Kundu_2021_WACV,Kundu_2021,datta_frontiers,datta2021training} and $N_{read}$ can be computed as
\begin{equation}
    N_{ac} = h_o*w_o*k^2*c_i*c_o 
\end{equation}
\begin{equation}
    N_{read} = k^2*c_i*c_o
\end{equation}

The energy values we have used to evaluate $E_{frontend}$ and $E_{backend}$ are presented in Table \ref{tab:energy_estimates}. While $E_{sens}$ and $e_{sense-to-tx}$ are obtained from our circuit simulations, $e_{tx}$ is obtained from \cite{wifi_energy}, and $e_{ac}$ and $e_{read}$ are obtained from \citep{kang2018memory}. Fig. \ref{fig:energy_comparison} shows the comparison of energy costs for standard vs P\si{^2}M-implemented spiking CNN models for the DVS datasets. In particular, P\si{^2}M can yield a backend energy reduction of up to $\sim2\times$ with the cost of $2\%$ increase in frontend energy only. This reduction primarily comes from the reduced energy consumption in the backend since we offload the compute of the first convolutional layer of the SNN. This layer consumes more than $50\%$ of the total backend energy since it involves expensive MAC operations (due to event accumulation before convolution computation) which consume ${\sim}32\times$ more energy compared to cheap accumulate operations \citep{horowitz20141} with 32-bit fixed point representation. Thus, the proposed neuromorphic-P\si{^2}M paradigm enables in-situ availability of the weight matrix within the array of DVS pixels (reducing the energy overhead associated with the transfer of weight matrix), while also significantly reducing energy-consumption of MAC operations by utilizing  massively parallel non-von-Neumann analog processing-in-pixel.  

\section{Conclusion}
In this work, we have proposed and implemented a novel in-pixel-in-memory processing paradigm for neuromorphic event-based sensors. To the best of our knowledge, this is the first proposal to enable massively parallel, energy-efficient non-von-Neumann analog processing-in-pixel for neuromorphic image sensors using novel weight-embedded pixels. Instead of generating event spikes based on the change in contrast of scenes, our proposed solutions can directly send the low-level output features of the convolutional neural network using a modified address event representation scheme. By leveraging advanced 3D integration technology, we can perform in-situ massively parallel charge-based analog spatio-temporal convolution across the pixel array. Moreover, we have incorporated the hardware (non-linearity, process variation, leakage) constraints of our analog computing elements as well as area consideration (limiting the maximum number of channels of the first neural network layer) into our algorithmic framework. Our P\si{^2}M-enabled spiking CNN model yields an accuracy of $88.36$\% on the IBM DVS128-Gesture dataset and achieved $\sim2\times$ backed energy reduction compared to the conventional system. 

\section*{Conflict of Interest Statement}
The authors declare that the research was conducted in the absence of any commercial or financial relationships that could be construed as a potential conflict of interest.

\section*{Author Contributions}
M.A.A.K. proposed the use of P\si{^2}M for the neuromorphic vision sensor and developed the corresponding circuit simulation framework. G.D. developed the baseline and P\si{^2}M-constrained algorithmic framework with the help of Z.W.  A.P.J. and A.R.J. proposed the idea of P\si{^2}M.  M.A.A.K. and G.D. wrote the majority of the paper.  A.R.J. and P.A.B. supervised the research and reviewed the manuscript extensively. All authors reviewed the manuscript.

\section*{Funding}
This work is funded in part by the DARPA HR00112190120 award.

\bibliographystyle{unsrtnat}
\bibliography{references}  






\end{document}